\documentclass[sigconf,screen=true]{acmart}

\usepackage{graphicx} %
\usepackage{geometry}
\usepackage{array}
\usepackage{enumitem}
\usepackage{tikz}
\usetikzlibrary{shapes.geometric}
\usetikzlibrary{arrows.meta, positioning}
\usetikzlibrary{fit, backgrounds}
\usepackage[acronym]{glossaries}

\newacronym{pgm}{PGM}{Probabilistic Graphical Model}
\newacronym{nn}{NN}{Neural Network}
\newacronym{gbdt}{GBDT}{Gradient Boosted Decision Trees}
\newacronym{llm}{LLM}{Large Language Model}
\newacronym{serp}{SERP}{Search Engine Results Page}
\newacronym{rctr}{RCTR}{Rank-Based CTR Model}
\newacronym{dctr}{DCTR}{Document-Based CTR Model}
\newacronym{rcm}{RCM}{Random Click Model}
\newacronym{pbm}{PBM}{Position-Based Model}
\newacronym{cm}{CM}{Cascade Model}
\newacronym{ubm}{UBM}{User Browsing Model}
\newacronym{dcm}{DCM}{Dependent Click Model}
\newacronym{dbn}{DBN}{Dynamic Bayesian Network Model}
\newacronym{ncm}{NCM}{Neural Click Model}
\newacronym{csm}{CSM}{Click Sequence Model}
\newacronym{pscm}{PSCM}{Partially Sequential Click Model}
\newacronym{tcm}{TCM}{Temporal Click Model}
\newacronym{thcm}{THCM}{Temporal Hidden Click Model}
\newacronym{emalgorithm}{EM}{Expectation-Maximization}
\newacronym{ips}{IPS}{Inverse Propensity Scoring}
\newacronym{mle}{MLE}{Maximum Likelihood Estimation}
\newacronym{ctr}{CTR}{Click-Through Rate}
\newacronym{ccm1}{CCM}{Carousel Click Model}
\newacronym{xpa}{XPA}{Cross-Positional Attentions}
\newacronym{gubm}{GUBM}{Grid-Based User Browsing Model}
\newacronym{ccm2}{CCM}{Click Chain Model}
\newacronym{cbcm}{CBCM}{Comparison-Based Click Model}
\newacronym{rbnn}{RBNN}{Rank-Biased Neural Network Model}
\newacronym{drlc}{DRLC}{Debiased Reinforcement Learning Click Model}
\newacronym{cacm}{CACM}{Context-Aware Click Model}
\newacronym{graphcm}{GraphCM}{Graph-Enhanced Click Model}
\newacronym{aicm}{AICM}{Adversarial Imitation Click Model}
\newacronym{fscm}{FSCM}{F-Shape Click Model}
\newacronym{tacm}{TACM}{Time-Aware Click Model}
\newacronym{pctm}{PCTM}{Probability Click Tracking Model}
\newacronym{bbm}{BBM}{Bayesian Browsing Model}
\newacronym{pcc}{PCC}{Post-Click Click Model}
\newacronym{gcm}{GCM}{General Click Model}
\newacronym{bss}{BSS}{Bayesian Sequential State Model}
\newacronym{fetcm}{FE-TCM}{Filter-Enhanced Transformer Click Model}
\newacronym{dcg}{DCG}{Discounted Cumulative Gain}
\newacronym{err}{ERR}{Expected Reciprocal Rank}
\newacronym{rbp}{RBP}{Rank-Biased Precision}
\newacronym{ope}{OPE}{Off-Policy Evaluation}
\newacronym{mnl}{MNL}{Multinomial Logit}

\copyrightyear{2026}
\acmYear{2026}
\setcopyright{cc}
\setcctype{by}
\acmConference[RecSys '26]{20th ACM Conference on Recommender Systems}{September 27-October 02, 2026}{Minneapolis, MN, USA}
\acmBooktitle{20th ACM Conference on Recommender Systems (RecSys '26), September 27-October 02, 2026, Minneapolis, MN, USA}
\acmDOI{10.1145/3773078.3831937}
\acmISBN{979-8-4007-2284-4/2026/09}

\title{From Click Modeling to Offline and Off-Policy Evaluation in Carousel Recommendation}

\begin{document}

\author{Jingwei Kang}
\affiliation{%
  \institution{University of Amsterdam}
  \city{Amsterdam}
  \country{The~Netherlands}
}
\email{j.kang@uva.nl}
\orcid{0009-0003-9283-4060}

\begin{abstract}
Carousel interfaces are widely used in modern recommendation systems. 
Unlike traditional interfaces that present a single ranked list, carousels simultaneously present several ranked lists to the user, as horizontally swipeable rows stacked on top of each other.
In this design, the rankings are closely tied to the two-dimensional layout.
Consequently, user behavior is shaped not only by item preference, but also by row organization, viewport constraints, and item context.
This tight coupling between ranking and presentation complicates the interpretation of user feedback, introducing new challenges for recommendation evaluation.

My PhD research aims to address these challenges by rethinking how carousel clicks are modeled and how carousel recommendation policies can be evaluated from logged interaction data.
So far, I have studied how users interact with carousel interfaces and developed a click model design framework that prioritizes mathematical relationships between observed variables over latent behavioral assumptions.
Building on these results, my ongoing work includes a project using discrete choice models to represent clicks as choices, alongside a project that develops carousel-specific offline metrics.
As a next step, I plan to develop off-policy evaluation methods that estimate the performance of recommendation policies from logged interactions.
Taken together, the expected contribution of my thesis is a connected body of work that links carousel click modeling with offline and off-policy evaluation, so that carousel recommendation policies can be improved more reliably.

\end{abstract}



\maketitle

\section{Introduction}

\begin{figure}[t]
    \centering
    \includegraphics[width=0.9\linewidth, alt={Netflix}]{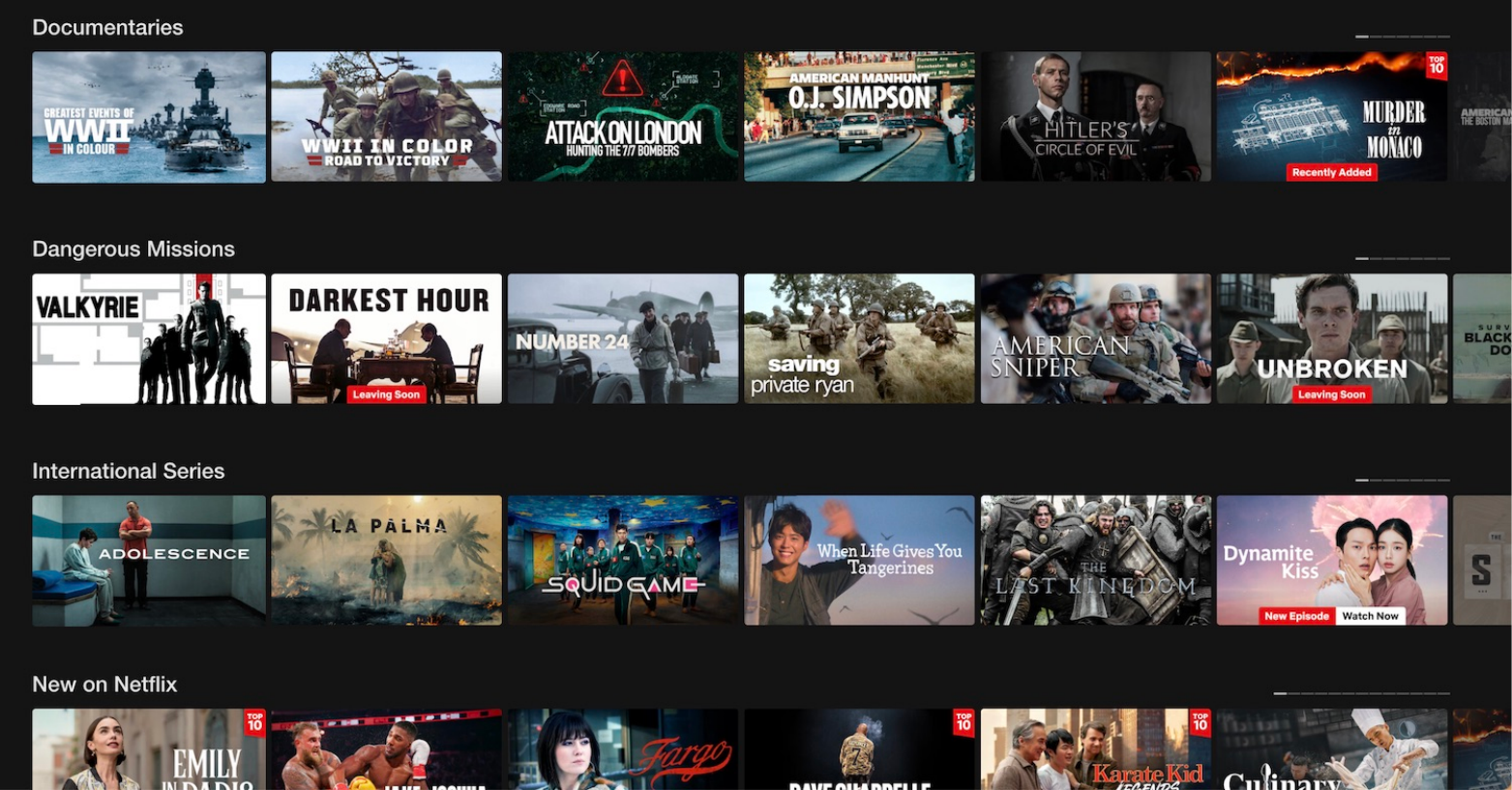}
    \caption{Example of movie recommendations by Netflix presented in a carousel interface.}
    \label{fig:carousel}
\end{figure}

In recent years, with the rise of streaming services such as Netflix and Spotify, the way recommendations are presented has changed significantly. 
Recommendations are no longer presented only as a single ranked list arranged in a particular layout, such as a search-results page~\cite{pmlr-v14-chapelle11a, Qin2010, 10.1145/2987380}, a vertical feed~\cite{Sun2023KuaiSAR, gao2022kuairand}, or a grid~\cite{10.1145/3077136.3080799, 10.1145/3308558.3313514, 10.1145/3726302.3730279}. 
Instead, carousel interfaces simultaneously present multiple horizontally swipeable ranked lists to the user in a vertical arrangement, as shown in Figure ~\ref{fig:carousel}.
Each list has a title or topic that describes its content, such as a specific type of movie or music, or a personalized category like \emph{Made for You} or \emph{Recently Played}. 
These interfaces have become increasingly popular for several reasons, in particular, their ability to support diverse user needs by presenting multiple collections of recommendations.

Offline and off-policy evaluation are key tools for developing recommendation systems from logged interaction data \cite{canamares2020offline,castells2022offline, 10.1145/3556536}. 
Although these methods usually treat clicks as rewards or indicators of user preference, in practice, click data is heavily biased by the user interface, the logging policy, and the resulting item exposure. 
As a result, offline and off-policy evaluation depend on a precise interpretation of what the observed feedback actually means. 
In the carousel recommendation, this is especially difficult. 
Because multiple ranked lists are displayed simultaneously, the system presents a structured two-dimensional page rather than a single list, shaping how users browse, compare, and choose.
Consequently, whether an item is clicked depends not only on user preferences, but also on its row assignment, its position within the row, and its surrounding context.

A common assumption in click modeling is that clicks can be decomposed into examination and relevance: users first examine an item, and conditional on examination, a click reflects the item's relevance or attractiveness~\cite{craswell2008experimental,chapelle2009dynamic}. 
My empirical work challenges this assumption in carousel interfaces. 
Using eye-tracking data, I find that the probability of a click, conditional on examination, remains position-dependent. 
This means that post-examination clicks cannot be treated as a position-invariant relevance signal.
Position and layout affect not only whether an item is examined, but also how, once examined, it is interpreted, compared with surrounding items, and ultimately clicked.

Together, these challenges motivate the \textbf{central focus of my thesis}: how carousel recommendation should be modeled and evaluated when the standard assumptions behind click modeling and evaluation framework no longer hold.

\section{Related Work}
\subsection{User behavior analysis}
A large body of user behavior analysis based on eye-tracking research \cite{Granka2008, 10.1145/1076034.1076063, 10.1145/1008992.1009079, 10.1145/3077136.3080799} has studied how users examine and interact with search and recommendation results, producing behavioral assumptions that are widely adopted in downstream model and metric design.
In single-list web search, this work has revealed a consistent examination pattern, known as the \emph{golden triangle}, sometimes also referred to as the F-pattern \cite{Granka2008}.
It is characterized by a triangle or F-shape that indicates users concentrate heavily on the top-left area of the page.
Eye tracking studies have motivated the \emph{examination hypothesis}, which assumes that a document must be examined before being clicked and that click probabilities are a product of examination and attractiveness probabilities~\cite{10.1145/1242572.1242643}. 
Similar F-shaped patterns have been observed in regular grid layouts, where each row contains a uniform number of items~\citep{10.1145/2959100.2959150, 10.1145/1935826.1935873}, while irregular grid layouts such as image search instead exhibit a clear middle-position bias, indicating that examination behavior is highly layout-dependent~\citep{10.1145/3077136.3080799}. 

For carousel interfaces, this empirical foundation is largely missing. 
The only eye-tracking dataset in this setting is RecGaze~\citep{10.1145/3726302.3730301}, introduced in this thesis (Section~\ref{sec:progress:recgaze}); to date it has been analyzed by one external follow-up focused on gaze transitions between carousels and between items within carousels~\citep{10.1145/3742413.3789166}, and by our own analysis of established behavioral assumptions in carousel interfaces~\citep{10.1145/3805712.3809657} (Section~\ref{sec:progress:sigir2026}).
As a result, prior to the work in this thesis, how users actually examine and click in carousel interfaces remained poorly understood.

\subsection{Click models}
Click models are probabilistic frameworks for interpreting user interactions, typically used to infer the relevance of real clicks~\citep{10.1145/3018661.3018699, 10.1145/3209978.3209986, 10.1145/3437963.3441794, 10.1145/3159652.3159732, 10.1145/2911451.2911537} and to predict future click probabilities to simulate user interactions~\citep{chuklin_markov_rijke_click_models_2015}. 
The field is dominated by single-list web search, where models are primarily built on \glspl{pgm} that encode hypotheses over latent user behavior, such as examination and attraction. 
Sequential examination models, including the \gls{cm}~\citep{10.1145/1341531.1341545}, \gls{ubm}~\citep{10.1145/1390334.1390392}, \gls{dcm}~\citep{10.1145/1498759.1498818}, and \gls{dbn}~\citep{10.1145/1526709.1526711}, assume users examine results from top to bottom; others, such as the \gls{thcm}~\citep{10.1145/2124295.2124334} and \gls{pscm}~\citep{10.1145/2766462.2767712}, allow non-sequential behaviors such as revisiting. 
Beyond single-list, click models have received some attention in grid layouts: \citet{10.1145/3209978.3209990} developed the \gls{gubm}, and \citet{10.1145/3442381.3450098} use \gls{nn} to learn click patterns in grids without restrictive assumptions.

For carousel interfaces, there is only one published click model, \gls{ccm1}~\citep{10.1145/3643709}: a cascade-style model in which users browse carousel topics until finding an attractive one and then browse items within it. 
However, the assumptions underlying this model do not fully account for two empirical findings in this thesis: how users browse carousel interfaces, and the fact that clicks remain position-dependent even conditional on examination.
Click modeling for carousel interfaces thus remains a largely open research area.

\subsection{Offline and off-policy evaluation}
Offline evaluation in search and recommendation has traditionally been based on single ranked lists, where users are often assumed to examine results sequentially from top to bottom.
Many list-based metrics can be understood as weighted-gain measures of the form \(M=\sum_i W_i R_i\)~\cite{10.1145/2505515.2507665}, where \(R_i\) is the relevance of the \(i\)-th result and \(W_i\) is the metric-specific weight at rank position \(i\).
\gls{dcg}~\cite{10.1145/345508.345545,10.1145/582415.582418} and \gls{rbp}~\cite{10.1145/1416950.1416952} use position-only weights, while \gls{err}~\cite{10.1145/1645953.1646033} uses cascade-dependent weights, in which $W_i$ depends on the relevance of documents ranked above position $i$.
For grid-based interfaces, \citet{10.1145/3308558.3313514} first proposes a grid-based metric of \gls{rbp}-style for the web images search, and then extends it to \gls{dcg} and \gls{err}.
Carousel interfaces require a further generalization of the weight term, since items are arranged in two dimensions and accessed through both horizontal and vertical navigation.
For this setting, previous work proposes N2DCG, an NDCG-style metric that incorporates the carousel structure into its position discount ~\cite{ferraridacrema2022offline,10.1145/3450614.3461680}.

However, when applied to logged feedback rather than ground-truth signals such as ratings, these metrics do not account for inherent exposure and selection biases of the logging policy~\cite{canamares2020offline,castells2022offline}.
An alternative view treats recommendation as an intervention or treatment: a recommender selects actions and observes outcomes only for those actions chosen by the logging policy~\citep{10.1609/aimag.v42i3.18141}.
This leads to \gls{ope}, where the goal is to estimate how a target recommendation policy would perform using logged interactions collected from a different logging policy.
Standard \gls{ope} methods include inverse propensity scoring and its variants~\citep{pmlr-v48-schnabel16,NIPS2015_39027dfa}, as well as doubly robust estimators~\citep{10.1214/14-STS500}.
In recommendation systems, actions are typically represented as ranked lists or slates, where multiple items and positions are jointly selected.
Consequently, \gls{ope} becomes substantially harder due to the combinatorial explosion of the action space. 
This challenge has motivated the development of slate-specific estimators under various user behavior assumptions, such as cascade \citep{10.1145/3488560.3498380, 10.1145/3394486.3403229}, independent \citep{10.1145/3219819.3220028,NIPS2017_5352696a,10.5555/3540261.3540541}, or context-adaptive \citep{10.1145/3580305.3599447}; a separate line avoids behavior assumptions through action-space abstraction for large action spaces \citep{10.1145/3589334.3645343,pmlr-v202-saito23b,pmlr-v162-saito22a}.

Carousel recommendation can be viewed as a multi-slate problem in which each slate itself is  structured, but it adds further interface constraints: users do not observe the full layout at once; instead, they navigate through a two-dimensional layout before clicking. 
The only existing study on carousel evaluation is limited to offline metric design; meanwhile, existing \gls{ope} work focuses mainly on single items or ranked lists.
This leaves a gap for carousel-specific \gls{ope} methods.

\section{Research Questions and Thesis Structure}
\label{sec:rq-structure}

My thesis is organized around two main research questions: 

\begin{itemize}
    \item \textbf{RQ1: How should carousel clicks be interpreted and modeled?}
    
    This question is addressed in the chapter \emph{Rethinking Carousel Clicks}. 
    This chapter studies carousel clicks from both an empirical and a modeling perspective. 
    It is divided into two sub-questions:
    \begin{itemize}
        \item \textbf{RQ1.1: How do users interact with carousel interfaces?}
        \item \textbf{RQ1.2: How can click models be designed for carousel interfaces?}
    \end{itemize}

    \item \textbf{RQ2: How should carousel recommendation policies be evaluated from logged interaction data?}
    
    This question is addressed in the chapter \emph{Evaluating Carousel Recommendation}. 
    This chapter studies carousel evaluation from two perspectives: metric design and off-policy evaluation.
    It is divided into two sub-questions:
    \begin{itemize}
        \item \textbf{RQ2.1: How can offline evaluation metrics be designed for carousel layouts?}
        \item \textbf{RQ2.2: How can off-policy evaluation be performed for carousel recommendation policies?}
    \end{itemize}
\end{itemize}

\begin{figure}[t]
\centering
\resizebox{0.9\linewidth}{!}{%
\begin{tikzpicture}[
    scale=0.9,
    transform shape,
    >=stealth,
    paper/.style={
        rectangle, rounded corners=3pt, draw, thick,
        text width=2.55cm, align=center,
        minimum height=0.9cm, inner sep=3pt, font=\scriptsize
    },
    published/.style={paper, fill=blue!15,   draw=blue!60},
    review/.style   ={paper, fill=orange!20, draw=orange!80},
    wip/.style      ={paper, fill=green!15,  draw=green!55, dashed},
    planned/.style  ={paper, fill=gray!12,   draw=gray!50, dotted},
    arrow/.style    ={->, thick, draw=black!70},
    rqbox/.style    ={draw=black!40, thick, rounded corners=8pt, dashed, inner sep=12pt}
]

\node[published] (ictir)    at (-2, 0)   {Click Model \\ Design Framework\\{\itshape ICTIR 2025}};
\node[published] (dataset)  at (2, 0)    {Eye-Tracking Dataset\\{\itshape SIGIR 2025}};

\node[wip]       (choice)   at (-2, -2)  {Choice-Based Click Model};
\node[published] (analysis) at (2, -2)   {Behavior Analysis\\{\itshape SIGIR 2026}};

\node[planned]   (ope)      at (-2, -4)  {Off-Policy Evaluation};
\node[review]    (n2dcg)    at (2, -4)   {Reformulation of N2DCG};

\begin{scope}[on background layer]
    \node[rqbox, fit=(ictir) (dataset) (choice) (analysis)] (rq1) {};
    \node[anchor=north west, xshift=2pt, yshift=-2pt, font=\small\bfseries, text=black!60] at (rq1.north west) {RQ1};

    \node[rqbox, fit=(ope) (n2dcg)] (rq2) {};
    \node[anchor=north west, xshift=2pt, yshift=-2pt, font=\small\bfseries, text=black!60] at (rq2.north west) {RQ2};
\end{scope}

\draw[arrow] (dataset)  -- (analysis);
\draw[arrow] (ictir)    -- (choice);
\draw[arrow] (analysis) -- (choice);
\draw[arrow] (analysis) -- (n2dcg);
\draw[arrow] (choice)   -- (ope);

\begin{scope}[shift={(0,-5.5)}, every node/.style={font=\scriptsize}]
    \node[published, text width=0.45cm, minimum height=0.28cm] at (-4.2,0) {};
    \node[anchor=west] at (-3.8,0) {Published};

    \node[review, text width=0.45cm, minimum height=0.28cm] at (-2.0,0) {};
    \node[anchor=west] at (-1.6,0) {Under review};

    \node[wip, text width=0.45cm, minimum height=0.28cm] at (0.4,0) {};
    \node[anchor=west] at (0.8,0) {Work in progress};

    \node[planned, text width=0.45cm, minimum height=0.28cm] at (3.2,0) {};
    \node[anchor=west] at (3.6,0) {Planned};
\end{scope}

\end{tikzpicture}
}
\caption{Structure of the thesis. The dashed bounding boxes categorize the papers by their corresponding RQs.}
\label{fig:thesis-structure}
\end{figure}
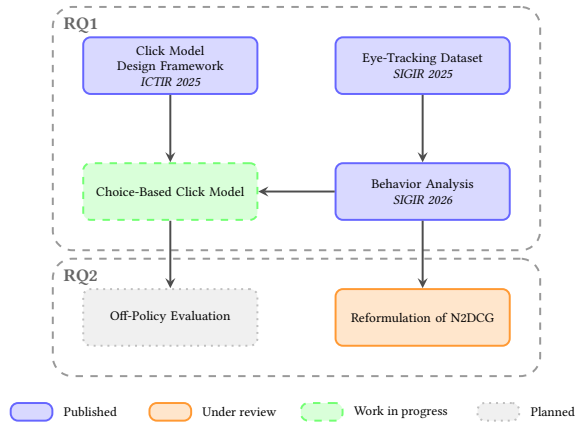

These questions are closely connected: reliable offline and off-policy evaluation requires a clear understanding of how users interact with carousel interfaces and what their clicks indicate.
Figure~\ref{fig:thesis-structure} shows the papers that make up this thesis and how they build on one another; we then relate each to the research questions above. RQ1.1 is addressed by the eye-tracking dataset (\emph{SIGIR 2025}) and the follow-up behavior analysis (\emph{SIGIR 2026}); RQ1.2 by the click model design framework (\emph{ICTIR 2025}) and the ongoing choice-based click model; RQ2.1 by the reformulation of N2DCG, currently under review; and RQ2.2 by the planned work on off-policy evaluation for carousel recommendation.

The arrows capture methodological dependencies: the dataset enables the behavior analysis, which informs both the choice model and the metric; the click-model framework also feeds the choice model; and the choice model provides the basis for the planned off-policy evaluation.
The N2DCG metric and the off-policy evaluation work are complementary components of the evaluation chapter: the former studies how to define carousel-aware offline metrics, while the latter studies how to estimate target-policy value from logged interactions.

\section{Research Progress}

\subsection{Chapter: Rethinking Carousel Clicks}

The first chapter of this thesis addresses RQ1, which asks how carousel clicks should be interpreted and modeled. 
The chapter consists of four papers: two empirical works addressing RQ1.1, and two modeling works addressing RQ1.2. 
The first two collect and analyze eye-tracking data on carousel interfaces to establish how users actually interact with them; the latter two translate these empirical findings to guide the design of click models.

\subsubsection{An eye-tracking dataset for carousel interfaces}
\label{sec:progress:recgaze}

\begin{figure}[t]
    \centering
    \includegraphics[width=\linewidth, alt={Netflix carousel interface}]{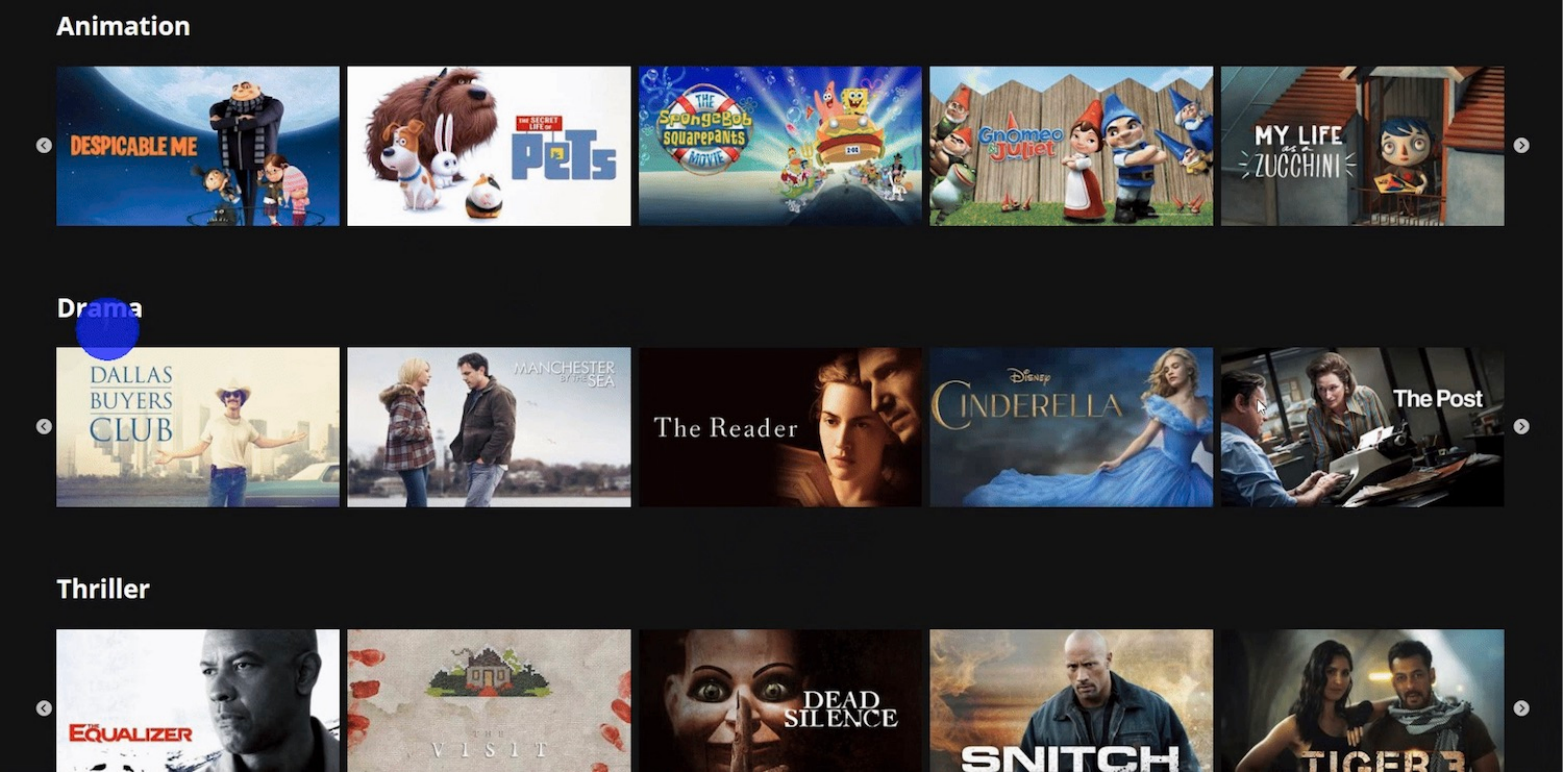}
    \caption{Design of a screen in the RecGaze eye-tracking study. The blue circle represents a user's current gaze point.}
    \label{fig:recgaze}
\end{figure}

To address RQ1.1, the first step is to collect the empirical data needed to study user behaviors directly. 
To this end, we constructed RecGaze~\citep{10.1145/3726302.3730301}, published at SIGIR 2025, the first dataset for carousel interfaces that combines eye-tracking data with clicks, cursor movements, and post-selection explanations.
The dataset was collected through a user study in which 87 participants completed three movie selection tasks, free browsing, semi-free browsing, and direct search, on 40 carousel screens, resulting in 3 {,}477 valid logged interactions. 
As shown in Figure~\ref{fig:recgaze}, each screen displayed 10 genre-based carousels of 15 movies, presented in an interface modeled after streaming services such as Netflix. 
An initial aggregate analysis of the gaze data suggested an F-shaped examination pattern on the first page, while deviations on subsequent pages motivated a more systematic analysis of whether this and other established behavioral assumptions hold in the carousel setting.

\subsubsection{Empirical analysis of carousel browsing behavior}
\label{sec:progress:sigir2026}

\begin{figure}[t]
  \centering
  \includegraphics[width=\linewidth]{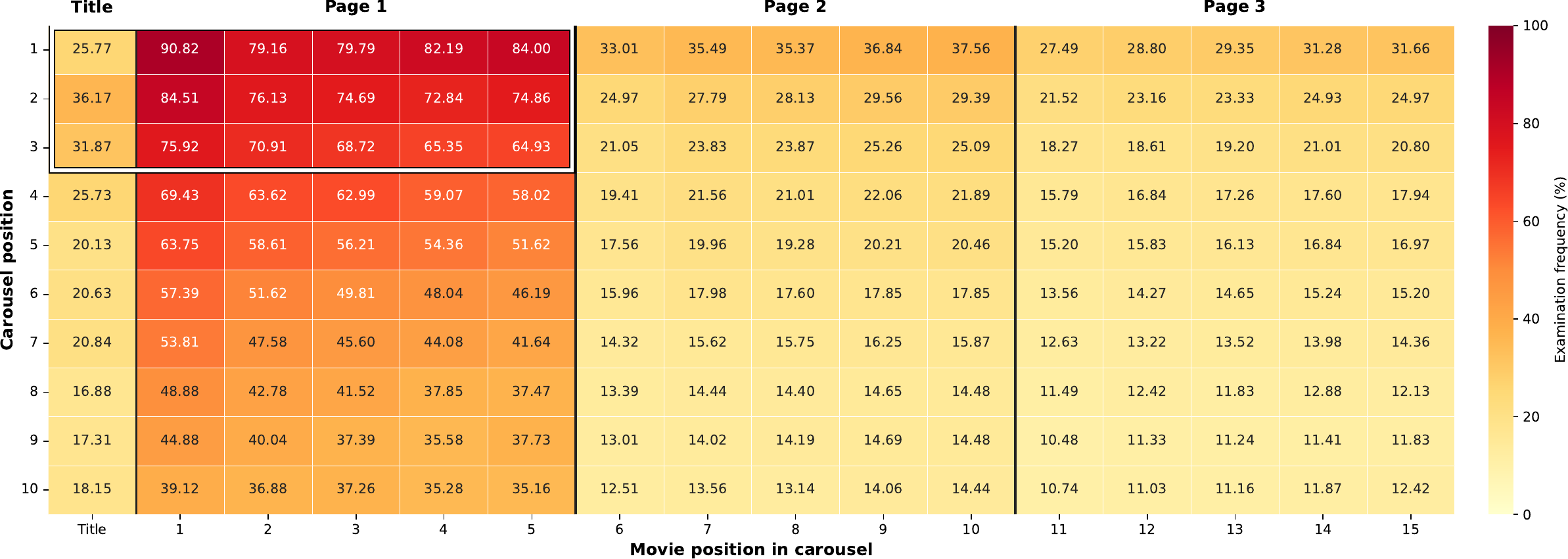}
  \caption{Empirical examination frequency heatmap across positions
  and pages. The number in each cell represents the frequency at that
  position; the white box marks the initially visible items and
  titles. The first page exhibits a dual-focus F-shaped pattern, while
  pages 2 and 3 exhibit mirrored F-shaped patterns; no global F-pattern
  spans the interface.}
  \label{fig:exam-prob}
\end{figure}

\begin{figure}[t]
\centering
\includegraphics[width=\linewidth, alt={Netflix carousel interface}]{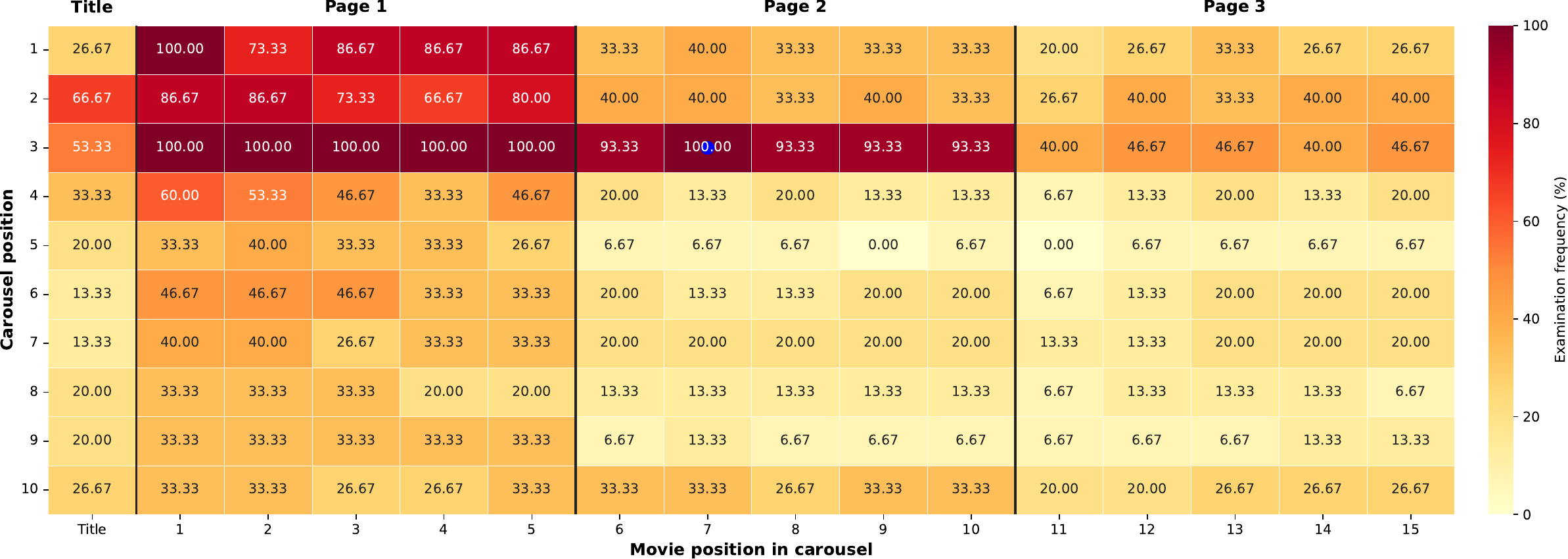}
\caption{Examination frequency given a click at position $(3,7)$ (marked by a blue point). Items above the clicked carousel on the first page and items preceding the click within the same carousel are examined with high probability, forming the L-pattern of carousel interfaces.}
\label{fig:L}
\end{figure}

\begin{figure}[t]
\centering
\includegraphics[width=\linewidth, alt={Netflix carousel interface}]{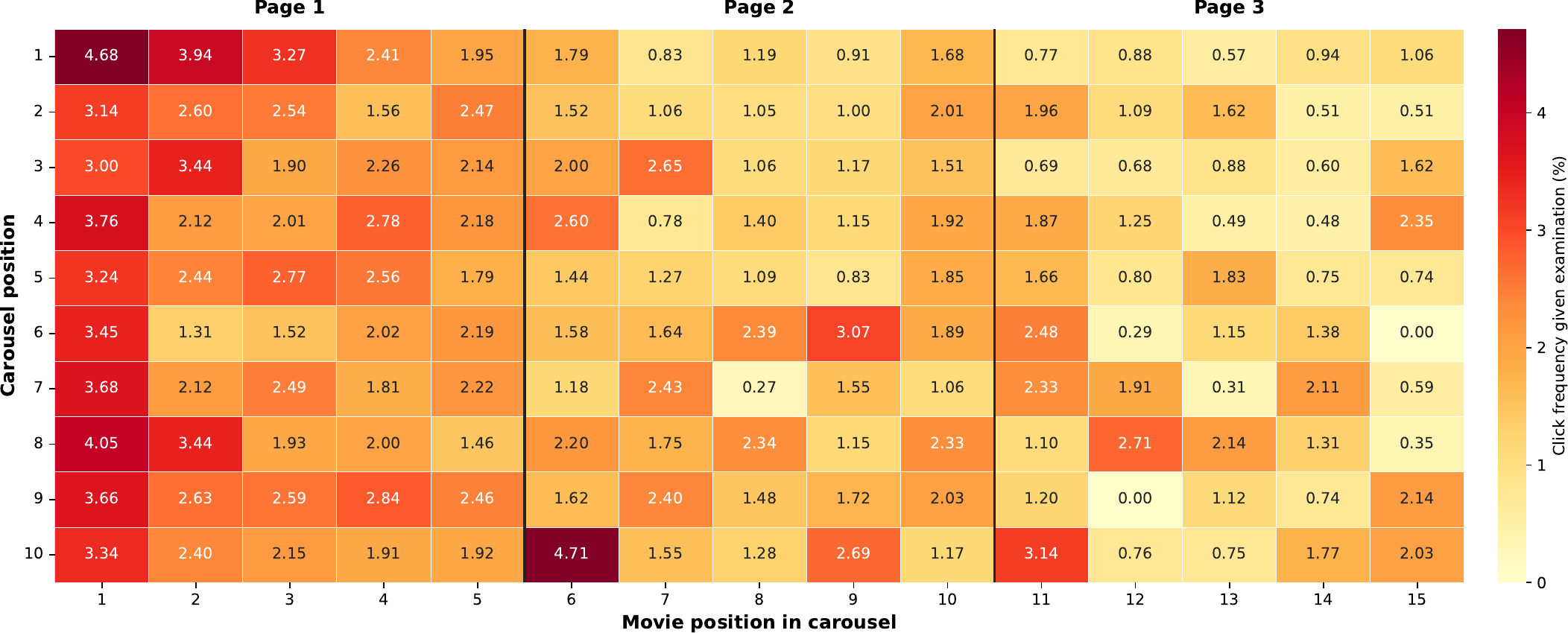}
\caption{Click frequency given examination per item position. The positions $(6,15)$ and $(9,12)$ are never clicked in the dataset, so the conditional click frequency is $0$.}
\label{fig:ctr}
\end{figure}

Building on RecGaze, this paper \cite{10.1145/3805712.3809657}, accepted at SIGIR 2026, systematically tests four behavioral assumptions that underlie existing carousel click models and evaluation metrics. 
In fact, the data refutes all four.
First, the classical F-pattern does not hold even on the first page, which instead exhibits a dual-focus structure with attention peaks in both top corners; after a horizontal swipe, the second and third pages exhibit a mirrored F-pattern rather than a continuation, so no global F-pattern spans the interface (Figure~\ref{fig:exam-prob}).
Second, examination conditioned on a click does not follow the assumed hierarchical, cascade-like pattern but instead forms a distinctive L-shape (Figure~\ref{fig:L}).
Third, conditional click probability given examination vary significantly across rows within the same column, despite the random vertical ordering of carousels in the study, refuting the examination hypothesis itself (Figure~\ref{fig:ctr}). 
Fourth, contrary to the assumption that users navigate carousels through their titles, the data shows that users typically examine the items first, and often without examining the corresponding title at all. 
Taken together, these findings indicate that the click models and evaluation metrics built on these assumptions require reconsideration.

\subsubsection{A theory-based framework for click model design}
A natural question follows from the empirical findings above: if standard behavioral assumptions do not hold in carousel interfaces, should they serve as the foundation for click model design at all?
This paper \cite{10.1145/3731120.3744585}, published at ICTIR 2025, reconsiders the conceptual foundations of click model design itself, grounding them in mathematical rather than behavioral assumptions.
We argue that the mathematical relationships captured between \emph{observed} variables should be the \emph{first concern} in click model design, rather than the latent variables or user behaviors they implicitly assume.
Building on this principle, we identify three key design choices that every click model makes, explicitly or implicitly: \textbf{global dependencies}, \textbf{sequentiality}, and \textbf{factorization}.
We then use these choices to construct a taxonomy that, for the first time, places single-list, grid, and carousel click models within a single comparative framework.
To demonstrate the value of the design framework, we derive a new carousel click model from these three choices, providing a starting point for the modeling work that follows in this chapter.

\subsubsection{A choice-based click model for carousel interfaces}
\label{sec:progress:choice-model}
This ongoing work follows the click model design framework mentioned above and models clicks directly as a discrete choice over the items a user has actually considered. 
The model captures click behavior as a two-stage process. 
First, a sequential browsing process generates a consideration set, which is initialized with the items visible at page load and expands through specific actions: swiping right within a row, scrolling down between rows, or stopping. 
These actions are bound by the physical constraints of the interface, preventing any skipping of pages within a row or skipping of rows altogether. 
Finally, a discrete choice over this generated set determines which item is clicked.
The click probability for item $i$ is
\begin{equation}
    P(\text{click } i \mid x) \;=\; \sum_{k:\, i \in C_k} P(C_k \mid x) \,\cdot\, \frac{\exp(u_i)}{\sum_{j \in C_k}\exp(u_j)},
\end{equation}
where $x$ denotes the context, $u_i \in \mathbb{R}$ is the utility of item $i$, $P(C_k \mid x)$ is the consideration-set distribution induced by the navigation process, and the fractional term is the within-set \gls{mnl} choice probability~\cite{mcfadden1974conditional,train2009discrete}.
The model captures two critical properties lacking in prior click models. 
First, an item's probability of being chosen depends on the other items it is jointly considered with, which examination-based models cannot represent.
Second, it explicitly models navigation in 2D carousel layouts under partial visibility, an aspect that standard 1D click models fail to address.
The methodological challenges, in this work, are parameterizing the consideration set construction process and the choice over that set, and estimating them from logged data in which only the final click is observed.

\subsection{Chapter: Evaluating Carousel Recommendation}

The second chapter of this thesis addresses RQ2, which asks how carousel recommendation policies should be evaluated from logged interaction data. 
The chapter consists of an offline metric for carousel layouts (RQ2.1), currently under review, and planned work on off-policy evaluation (RQ2.2).

\subsubsection{A behavior-grounded reformulation of N2DCG}
To address RQ2.1, we revisit the only existing carousel evaluation metric, N2DCG~\cite{10.1145/3450614.3461680,ferraridacrema2022offline}, adapted from NDCG to two-dimensional carousel layouts.
This paper, currently under review, identifies two substantial limitations of the original formulation. 
First, its ideal ranking, which is used for normalization, violates the categorical constraints of the carousel interface. 
Second, its discount function inherits assumptions from single-list web search that our empirical analysis in Section~\ref{sec:progress:sigir2026} shows do not hold in carousel interfaces.
We propose a reformulation that addresses both limitations.
Empirically, the reformulated metric tracks user behavior on real-world eye-tracking data more faithfully than the original N2DCG. 
Furthermore, it better predicts the outcomes of simulated carousel layout comparisons based on empirical examination frequencies.

\subsubsection{Off-policy evaluation for carousel recommendation}
To address RQ2.2, the planned work aims to develop \gls{ope} methods for carousel recommendation policies, filling the gap identified in the related work above.
Standard slate \gls{ope} estimators assume a single, fully observable ranking: all items in the slate are simultaneously on display at presentation time, and the modeled structure concerns only the order in which the user examines them.
Carousel interfaces do not fit this setting: the layout contains multiple rankings stacked vertically, and at any moment only a portion of each visible ranking is on screen, with further items accessed through horizontal swiping and further rankings through vertical scrolling.
The planned work therefore explores how existing slate \gls{ope} methods can be adapted to this multi-ranking, partially visible setting, along two possible routes.
One route reduces the two-dimensional layout to a one-dimensional slate: building on the discrete choice model of Section~\ref{sec:progress:choice-model}, we can estimate the most likely consideration set and collapse the layout into a single effective ranking, to which standard slate estimators directly apply.
The other route retains the two-dimensional structure but organizes it hierarchically, representing the layout as a slate of sub-slates, so that cross-ranking and within-ranking behavior are modeled as a two-stage process.

\section{Conclusion}

This thesis investigates how carousel recommendations should be modeled and evaluated. 
This exploration is motivated by the fact that existing click models and evaluation methods for single-list web search fail to transfer to carousel layouts.
Empirically, work using the RecGaze eye-tracking dataset and follow-up analysis has established a behavioral foundation for user interaction with carousels. 
Theoretically, this research introduces a click model design framework that prioritizes observable variables over latent behaviors.
Building on these foundations, my ongoing and planned work develops a choice-based click model for carousels and pursues two complementary directions for off-policy evaluation, alongside a behavior-grounded reformulation of N2DCG that is currently under review.

Together, these contributions aim to provide the modeling and evaluation tools that are currently lacking in the carousel recommendation.
At this stage of the thesis, feedback from the doctoral symposium on the design of the two planned off-policy evaluation routes and on the broader evaluation methodology for carousel recommendation would be highly valuable.

\begin{acks}
I would like to thank my supervisors, Maarten de Rijke and Harrie Oosterhuis, for their guidance, feedback, and support.
I am also grateful to my collaborator, Santiago de Leon-Martinez, for designing the eye-tracking study.
My PhD is supported by the Dutch Research Council (NWO) under grant \href{https://www.nwo.nl/en/projects/kich3ltp20006}{KICH3.LTP.20.006}.
All content represents the opinion of the authors, which is not necessarily shared or endorsed by their respective employers and/or sponsors.
\end{acks}

\newpage
\bibliographystyle{ACM-Reference-Format}
\balance
\bibliography{references}

\end{document}